\documentclass[conference]{IEEEtran}
\usepackage{xcolor}
\usepackage{diagbox}

\usepackage{balance}

\usepackage{cite}

\ifCLASSINFOpdf
\usepackage{graphicx,epstopdf}
\else
\fi
\usepackage{amsmath}
\usepackage{amssymb} 

\usepackage{booktabs} 
\usepackage{array}

\ifCLASSOPTIONcompsoc
  \usepackage[caption=false,font=normalsize,labelfont=sf,textfont=sf]{subfig}
\else
  \usepackage[caption=false,font=footnotesize]{subfig}
\fi
\begin{document}
%
\title{Radar Human Motion Recognition Using Motion States and Two-Way Classifications}

\author{\IEEEauthorblockN{Moeness G. Amin and Ronny G. Guendel}
\IEEEauthorblockA{Center for Advanced Communications, Villanova University, USA (moeness.amin,~rguendel@villanova.edu)}
}


%


\maketitle

\begin{abstract}
We perform classification of activities of daily living (ADL) using a  Frequency-Modulated Continuous Waveform (FMCW) radar. In particular, we consider contiguous motions that are inseparable in time. Both the micro-Doppler signature and range-map are used to determine transitions from translation (walking) to in-place motions and vice versa, as well as to provide motion onset and the offset times. The possible classes of activities post and prior to the translation motion can be separately handled by forward and background classifiers. The paper describes ADL in terms of states and transitioning actions, and sets a framework to deal with separable and inseparable contiguous motions. It is shown that considering only the physically possible classes of motions stemming from the current motion state improves classification rates compared to incorporating all ADL for any given time.

\end{abstract}


%
\IEEEpeerreviewmaketitle

\section{Introduction\label{sec:intro}}

Monitoring of Activities of Daily Living (ADL) finds applications in assisted living and "aging-in-place" and provides means for living independently 
\cite{1_aminSP, 2_amin2017radar, 8_4801689, 11_bijan2009, 10_5404249, 9_7348882}. 
Among ADL, falls are considered as an abnormal activity that should be accurately detected and classified with high sensitivity and specificity 
\cite{12_Hong2011, 13_Su2015, 14_Wu2015, LeKernec8746868}. 
Other daily activities can indicate, in their variants over times, changes in lifestyle as well as the state of physical, cognitive, and psychological health of a human being. In addition to ADL, RF-based gesture recognition using hands and arms has become an important contactless technology for Man-Machine-Interface (MMI) 
\cite {kim2016hand, wang2016interacting, skaria2019hand, maminzz, zhang2016dynamic}. 
RF-based vital sign and gait monitoring has vast medical applications and offers essential diagnostic barometers for many health problems 
\cite{2_amin2017radar, 3_7842648, Seifert8613848, 5_7944373, 6_seifert2019RadarConf}.  
In this paper, we focus on the classification of ADL, dealing with motions as contiguous activities which occur in certain norms and sequences consistent with the human ethogram \cite{humanEthogram}. The latter is a catalog of motion behaviors. This is in lieu of the commonly assumed isolated motions with no implication to previous and follow-on actions.

Human daily activities can be categorized into translation and in-place motions. Whereas the former describes crawling and gait articulations, the latter is primarily associated with motions that do not exhibit considerable changes in range. In-place motions include sitting, standing, kneeling, and bending, each is performed without any stride. Fall can be considered a translation or in-place motion, depending on whether it is a “progressive” or “heart attack” type.

We use Two-Dimensional (2-D) Principal Component Analysis (PCA), a data-driven feature learning technique, followed by k-NN classifier. The 2-D~PCA has shown to be very effective in human motion classifications 
\cite{baris:DataCubeProcessing8691492, park2016microdoppler_bc, maminzz}. 
It outperforms hand-crafted based classifications and offers competitive results to convolution and deep neural networks. Both the target micro-Doppler signature, provided by the spectrograms of the radar signal returns, and the target range-map are input to the 2-D~PCA. It is noted that the essence of this paper contribution is not to devise new classifier but rather address the contiguity issue of human motions and exploit the nature of human activities which limit the possible contingent motions stemming from the present one.

In this paper, we use a Frequency-Modulated Continuous Waveform (FMCW) radar with range and Doppler resolution capabilities. We first separate translation and in-place motions using the Radon transform which is applied to the range-map to detect the piece-wise linear behavior of range with respect to slow time of a moving target. Accordingly, horizontal lines correspond to in-place motions, whereas translation motions are manifested by lines with non-zero slopes. The Radon transform, instead of a motion tracker \cite{book:principleOfModernRadar}, can simply reveal the transitions from translation to in-place motion and visa versa by capturing the "breaking" points, or time instants of changing slopes. Over the in-place motion time segments, an energy detector is applied to possibly determine the onset and offset times of each motion.

We consider classifying consecutive motions incorporating the ethogram sequence of human motion articulations. For example, a person does not walk from a bending down posture unless first there is a rebounding up from bending fulfilled, and the person does not immediately sit down after walking unless there is a pause where there is an adjustment of posture to enable sitting. 
In this paper, we cast the ADL as states connected by motion actions. Each of the defined states of walking, sitting, standing, and laying has possible prior and post actions, according to the human ethogram. 
The elimination of impossible motions leads to varying the classes of motions considered at any given time. This, in turn, (a) alters the size of the classifier confusion matrix, instead of using a fixed matrix dimension that is associated with accounting for all motion classes in the classifier decision. (b) allows the use of different classifiers at different motion states.

The paper is organized as follows. Section~\ref{sec:2theory} introduces the experimental setup for data collection with a FMCW radar. Section~\ref{sec:3hsr} describes the proposed algorithm, including forward and reverse time motion classification and state representation. Section~\ref{sec:4radon} illustrates the intra- and inter-class separation technique, and Section~\ref{sec:experimentalResults} provides experimental results based on two examples. The conclusions and remarks are given in Sec.~\ref{sec:conc}. 

\section{Radar System and Data Analysis\label{sec:2theory}}

\subsection{\label{subsec:RadarModel}Radar Model}

The data collection was performed using SDRKIT 2500B, which is developed by Ancortek, Inc. The FMCW radar operates with a center frequency of $25~GHz$ and a bandwidth of $2~GHz$. The pulse repetition interval (PRI) is  $1~ms$, and the range resolution is ~$7.5~cm$.   The transmitted signal is,
%
\begin{equation}
    S_{Tx}(t) = A_T \cdot cos[2\pi (f_c t +\frac{1}{2}\alpha t^2)]
\end{equation}
\noindent where  $\alpha$ is  the chirp rate given by $B/T$, and $B$ indicates the bandwidth. The received signal is,
\begin{equation}
    S_{Rx}(t) = A_R \cdot  cos[2\pi (f_c (t-\tau) +\alpha (\frac{1}{2} t^2 -\tau\cdot t) + f_D \cdot t)]
\end{equation}
\noindent where $\tau$ is the two-way travel time, $f_D$ is the Doppler shift assuming constant velocity target, and $A_R$ is the signal amplitude. The complex baseband signal is expressed in terms of the in-phase and quadrature components as, 
\begin{equation}
s(t) = I(t) + jQ(t) = Ae^{\psi (t)}
\end{equation}
\noindent where ($\psi(t)$)   is the signal phase \cite{baris:DataCubeProcessing8691492, art:fmcwRadarModel}. This signal is used in the follow-on analysis.

\subsection{Range-map Computation \label{subsec:rangemap}}

For the computation of the target range profile, the matched filtered of the radar return is represented by a two-dimensional matrix, $s(n,m)$. The Discrete Fourier Transform (DFT) applied to each column, corresponding to one $PRI$, provides the target range information. The range-map, $R(p,m)$, is generated by incorporating consecutive $PRI's$, and it is given by, 
\begin{equation} \label{eq:STFT:RM}
R(p,m) = \frac{1}{N} \sum_{n=0}^{N-1} s(n,m) exp(-j 2 \pi \frac{p n}{N})
\end{equation}
\noindent where $p = 0, . . . ,N - 1$,  $N$ is the number of samples, or range bins, in one $PRI$,  and $m = 0,...,M-1$, where $M$ represents the total number of $PRI's$ considered. In our data collection experiments, we set N and M to be $512$ and $8,000$ (e.g. for eight seconds of data), respectively.

\subsection{Micro-Doppler Signature}

To obtain the target micro-Doppler signature, we first sum the data over the range bins of interest as, 
\begin{equation} \label{eq:rangebinsum}
V(m) = \sum_{r =r_1}^{r_2} R(r,m)
\end{equation}
\noindent where $r_1$ and $r_2$ are the minimum and maximum range bins considered, set to  $10$  and $128$, respectively. This corresponds to a range swath from $0.75~m$ to $9.6~m$ \cite{fmcwRadar_8443507}. The short-time Fourier transform (STFT) is then applied to $V(m)$, and its magnitude square, i.e., spectrogram, is computed to yield the micro-Doppler signature, $MD(n,k)$,
\begin{equation} \label{eq:MDequation}
MD(n,k) =  \left|	\sum_{m=0}^{L-1}w(m)V(n-m)exp(-j 2 \pi \frac{m k}{L}) \right| ^2
\end{equation}
\noindent A Hanning window $w(m)$ of size $L = 128$ is applied to reduce the sidelobes \cite{art:hanningWindow467238}. In Eq.~\ref{eq:MDequation}, a shift operator ($D$) of $8$ samples, which corresponds to $94~\%$ window overlapping, is used. The spectrogram is resized with $128$~samples for Doppler scaling and $32~samples = 1~s$ in slow-time. The same resizing process is applied for range-map images.

\subsection{Feature Extraction and Classification\label{subsec:featureExtr}}

The Two-Dimensional Principal Component Analysis (2-D~PCA) is used is used for feature extraction, followed by the k-Nearest-Neighbours (k-NN) classifier. The covariance matrix $ H $ is computed as, 
\begin{equation}\label{ImageCovMatr}
H = \frac{1}{I} \displaystyle\sum_{i=1}^{I} (X^{(i)}-\bar{X})^T\cdot(X^{(i)} -\bar{X})
\end{equation}
\noindent
where $\bar{X} \in \mathbb{R}^{\eta \times \eta}$  is the mean image and $I$ is the total number of images in the training data. In the above equitation, $X^{(i)} \in \mathbb{R}^{\eta \times \eta}$ is the $i$-th micro-Doppler or range-map image, computed form $MD(n,k)$ and $R(p,m)$, respectively. From the eigendecomposition of $ H $, the eigenvalues ($\lambda_i$) and eigenvectors ($\nu_i$)  are extracted, such that $ J(\Phi)=\Phi^T H \Phi $. The eigenvectors, corresponding to the  $d$ largest eigenvalues form the matrix $\Phi = [\nu_1, \nu_2, ..., \nu_d$]. In our work, the default setting was $d_{MD}= 14$ and $d_{RM}=4$ for the principal eigenvectors, used for micro-Doppler and range-map classification, respectively. The individual training images $X^{(i)}$ are projected onto  the $d$-dimensional subspace matrix to compute the principal component matrix,  $Y = X\Phi$. The micro-Doppler image, $Y_{MD}$  is of  dimension $\mathbb{R}^{\eta \times d_{MD}}$, and the range-map image, $Y_{RM}$ is of dimension $\mathbb{R}^{\eta \times d_{RM}}$. 

The individual test images are projected using the same procedure to provide the feature matrix $Y_{MD}(Test)$ and $Y_{RM}(Test)$. 
The k-NN classifier operates on the fused vectorized and concatenated micro-Doppler and range-map feature vectors \cite{baris:FusionPCA8835840}.

\section{Human Motion State Representation\label{sec:3hsr}}

\begin{figure}[htbp]
\centering
\includegraphics[width=\linewidth]{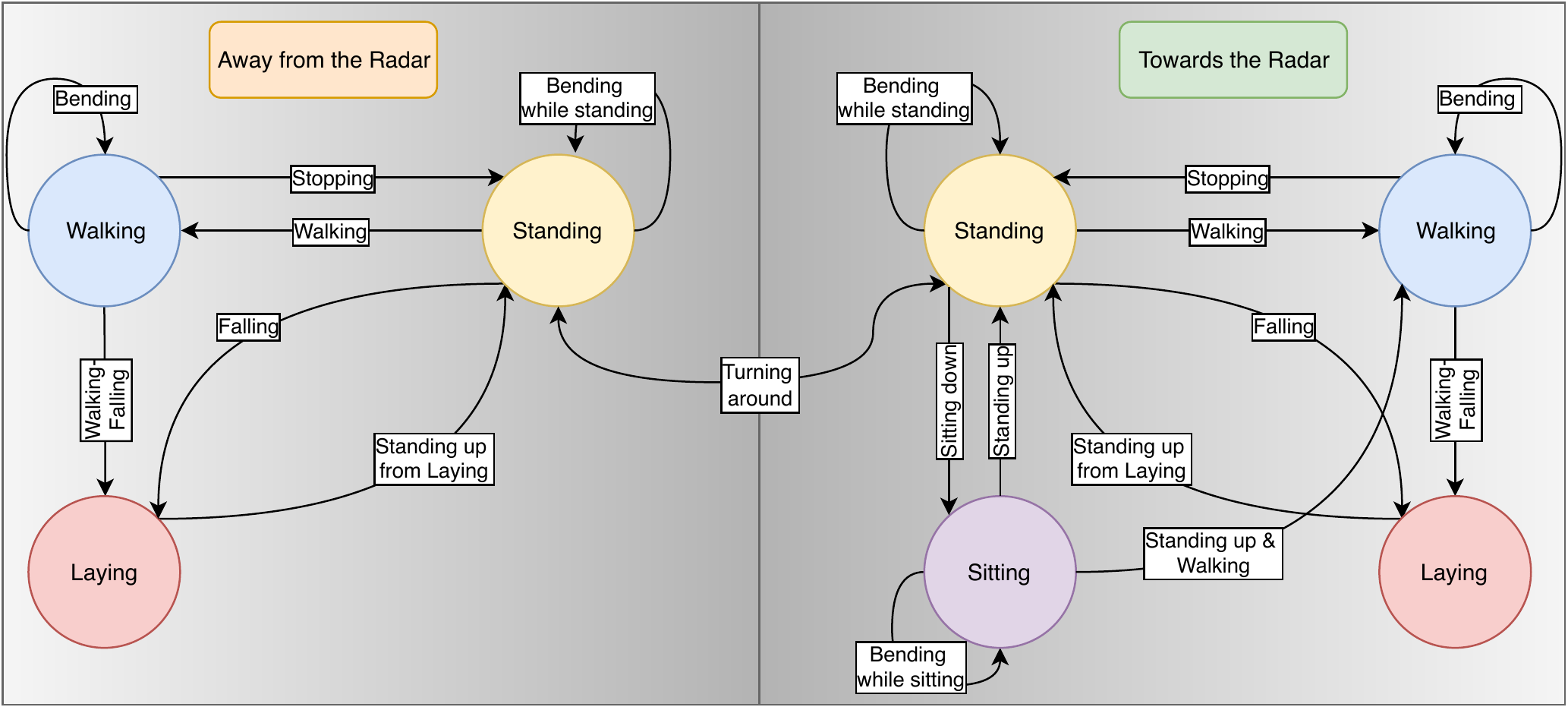}
\caption{State diagram for forward time motion detection. \label{fig:flowgraph_fwd}}
\end{figure}

In this section, we cast human activities as states, namely, walking, standing, sitting, and laying. A change, or a transition, from a state to another is performed through an action, or an activity. This is shown in the state diagram in Fig.~\ref{fig:flowgraph_fwd}.
%

\subsection{Forward Time Motion Sequence for "Walking State"\label{subsec:31hsr}}
The person changes from a “walking state” (WS) to a “laying state” (LS) through a falling down action, and transition to a “standing state” (StS) through a stopping action. It can also transition to itself through a bending down and up action. In the latter case, there is not sufficient time separation between walking followed by bending down and between rebounding from bending and walking in order to be able to declare a (StS) in each case. The classifier must then discriminate between these three actions which are represented by arrows emanating from the WS in Fig.~\ref{fig:flowgraph_fwd}. 

\subsection{Reverse Time Motion Sequence for “Walking State”\label{subsec:32hsr}}
The reverse time sequence of actions considering walking is not entirely reciprocal to the forward sequence. For example walking cannot be preceded by falling but can be followed by it. Also, the WS can be directly reached from the SiS, through standing up action, but cannot be followed by it. This is because a person needs to exhibit short time duration of standing between walking and sitting down for body adjustment which implies that the SiS is preceded by the StS and not by the WS. However, standing up from sitting followed up by walking can be merged with no time in between to declare a StS. It is noted that such merging is only possible if standing up in the direction of the follow-on walking motion, otherwise the person would need to turn around after standing up and walk in the opposite direction which gives rise to a short time interval where the person is in the StS.  The same is true for bending. Accordingly, a classifier needs to consider only three actions prior to walking which are indicated by the arrows entering the walking state in Fig.~\ref{fig:flowgraph_fwd}. 

\subsection{Forward and Reverse Time Motion Sequence for “Sitting State”\label{subsec:33hsr}}
In the forward time motion sequence, a person can transition from the SiS to itself though bending action, as shown in Fig.~\ref{fig:flowgraph_fwd}. It can change to the StS by a standing up action or to a WS, as discussed above.  So a classifier applied in the SiS needs to consider only three classes. For the reverse time motion sequence, a change into the SiS can be performed from the StS only. This is also shown in Fig.~\ref{fig:flowgraph_fwd}. 

\subsection{Forward and Reverse Time Motion Sequence for “Standing State”\label{subsec:34hsr}}
In the forward time motion sequence, a person can transition from the StS to itself though a bending action, as depicted in Fig.~\ref{fig:flowgraph_fwd}. It can change to the WS, SiS, or LS. In the reverse time motion sequence, changing into the StS can be from the SiS through standing up, from the WS through stopping, from a LS through standing up from falling, and transition from itself through bending. In this regard, the StS is associated with the highest number of motion classes, or actions, in the forward and reverse time directions.

\subsection{Forward and Reverse Time Motion Sequence for “Laying State”\label{subsec:35hsr}}
In the forward time motion sequence, a person can change from the LS to the StS through a standing up motion, which is the only possible action. The person can change into the LS from the StS through falling and from the WS also through falling. This is shown in the diagram of Fig.~\ref{fig:flowgraph_fwd}.  

The diagram in Fig.~\ref{fig:flowgraph_fwd} suggests two possible classifiers, each is applied to the motion actions transitioning in and out of a state. Once a state is detected, then the two associated classifiers for the in and out transitions can be applied to infer the previous and follow-on motions, respectively. It should be emphasized that each state can apply a different classifier than the rest.

\subsection{Motions Towards and Away from the Radar\label{subsec:36hsr}}
To account for the possibilities that the motion actions can be performed towards and away from the radar, we generalize the state diagram in Fig.~\ref{fig:flowgraph_fwd} to consist of two state groups, namely, Group-T for toward radar motions and Group-A for away radar motions. Each group represents one motion direction. A person can transition across the two groups by the means of turning around while standing.

\section{Radon Transform for Range-Map Processing\label{sec:4radon}}

The Radon transform \cite{29_wininger2013basis} is considered an effective tool in detecting dominant contour schemes in images, especially in medical image processing. In the underlying problem, the goal is to apply the Radon transform to detect pertinent line structures. In employing the Radon transform, we recognize that horizontal lines in the range-map correspond to in-place motions with no noticeable range swath, whereas lines with non-zero inclinations represent continuous changes in range gates stemming from motion translations, such as constant speed walking. It is noted that acceleration or deceleration gives rise to curvy signatures in the range-map in which case, the Hough transform can be applied \cite{32_7944316}.

\subsection{Application of the Radon Transform\label{subsec:radonTrans}}

We consider the range-map as an image, $ B(m,n)$, with each sample converted to a decibel absolute value. The radar data collected produces an image of size $ M \times N = 256 \times 12,000 $, where $ M $ and $ N $ represent the number of range bins (rows) and slow-times (columns), respectively.

\begin{figure}[hbt!]
    \centering
 \subfloat[\label{subfig:RM4}]{%
      \includegraphics[width=0.49\linewidth]{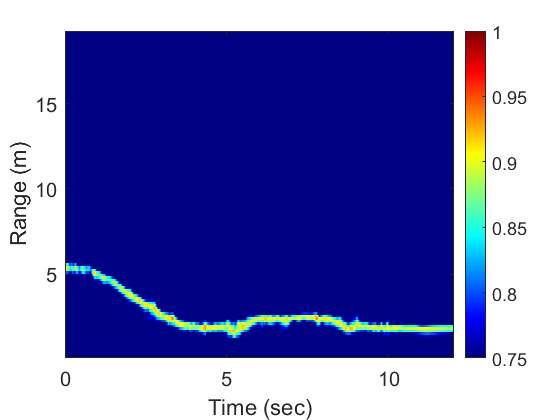}}
    \hfill
  \subfloat[\label{subfig:RM_Lines}]{%
        \includegraphics[width=0.49\linewidth]{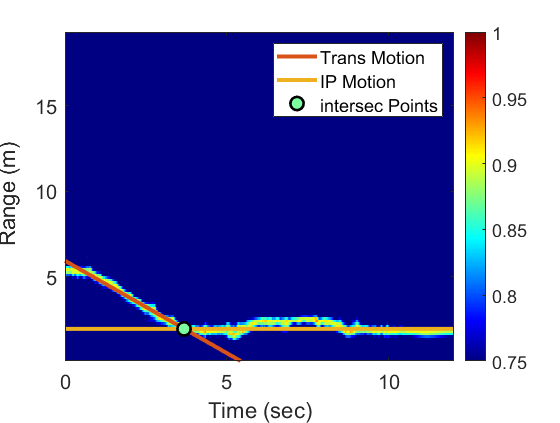}}
    \\
  \subfloat[\label{subfig:MD_Separation}]{%
        \includegraphics[width=0.49\linewidth]{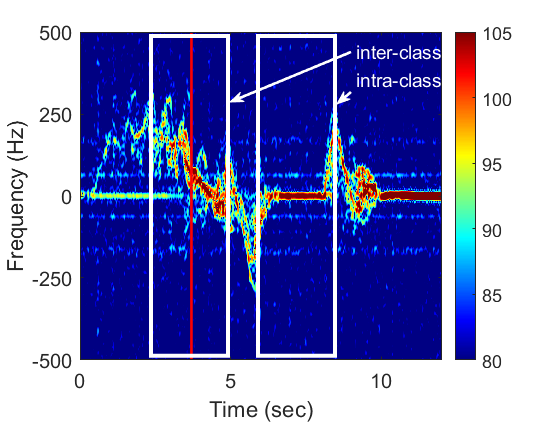}}
    \hfill
  \subfloat[\label{subfig:PBC3}]{%
        \includegraphics[width=0.49\linewidth]{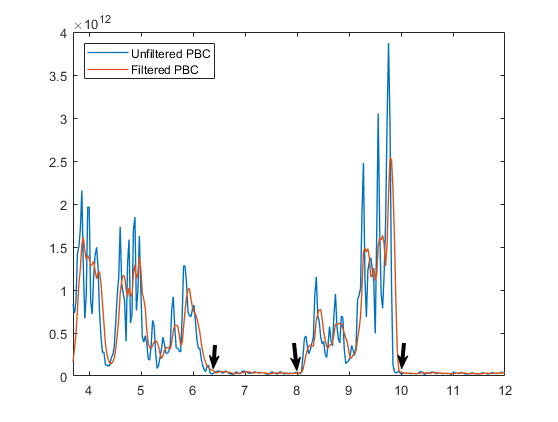}}
  \caption{The image set shows (a) range-map for the motion sequence, (b) range lines and intersection point, (c)  micro-Doppler image and (d) computed energy of the in-place segment.}
  \label{fig:RM_MD_PBC} 
\end{figure}

Fig.~\ref{subfig:RM4} shows an example of the range-map where the person begins walking, then assumes two consecutive in-place motions, namely, sitting down and standing up. The range-map image is resized, filtered, and thresholded.
Image resizing is performed by uniform sub-sampling over slow-time to produce a smaller image size, referred to as $ RM_{ds} $,  of dimension $ 128 \times 384 $. To improve the end result, the down-sampled image $ RM_{ds} $ is filtered with a ($ 3\times 3 $) smoothing kernel of unit value coefficients.
\noindent Fig.~\ref{subfig:RM_Lines} shows the two lines detected by applying  the Radon transform on Fig.~\ref{subfig:RM4} as well as their intersection point.

\subsection{Power Burst Curve (PBC)\label{sec:PBC}}
\noindent To determine whether there is one or a sequence (multiple) of in-place motions, we examine the micro-Doppler signatures over the in-place motion interval. The spectrogram of the motions in the above example is shown in Fig.~\ref{subfig:MD_Separation}. For separating the two consecutive in-place motions, namely sitting down and standing up, we measure the rise and fall of the signal energy in $MD(n,k)$ over slow-time which is known as the Power Burst Curve (PBC) \cite{32_7944316, 33_6889337}.

The selected frequency bands for power computation are bounded by \(K_{P1} = 20~Hz\) and \(K_{P2} = 270~Hz\)  for  positive-Doppler frequencies and  by 
\(K_{N2} =  - 20~Hz\) and \(K_{N1} =  - 270~Hz\) for negative Doppler frequencies. The PBC for the combined frequency bands is given by, 
\begin{equation}
    \label{eq:PBC}
    \begin{split}
    PC(n) = \sum\limits_{k_1 = {K_{P1}}}^{K_{P2}} {{{\left| {MD(n,{k_1})} \right|}^2}}  + \\\sum\limits_{{k_2} = {K_{N1}}}^{{K_{N2}}} {{{\left| {MD(n,{k_2})} \right|}^2}}, n = 1,2,\dots,N
    \end{split}
\end{equation}
The computation of the above equation results in a fluctuating power curve stemming from intricate micro-Doppler signatures of human motions. Such fluctuation could mistakenly define wrong event boundaries.
%
%
To mitigate the above problem, we apply a moving average filter with an extent of $w=5$ samples. The filtered PBC, shown in Fig.~\ref{subfig:PBC3}, is used to determine the onset and offset times of each activity. The threshold has been found empirically as $3\%$ over the minima as, $PC_{f_{min}}+ 0.03 \cdot(PC_{f_{max}}-PC_{f_{min}})$.

In constructing the state diagram in Sec.~\ref{sec:3hsr}, we included human motions which could easily merge. This represents a challenge to the PBC to separate motion events. So, we rely on the breaking point generated by the Radon transform to indicate in-place motion occurrence and use a window around it to capture the corresponding action.

\section{Experimental Results \label{sec:experimentalResults}}
The example in Fig.~\ref{fig:SequenceWlkFalSta} shows a falling incorporated with a prior walking, followed by a laying on the floor period, followed by the person standing up from falling to reach the StS and, finally,  pursuing a walking motion.

\subsection{Post-Walking Motion Classification} 
\begin{figure*}[htb]
\centering
\includegraphics[width=\linewidth]{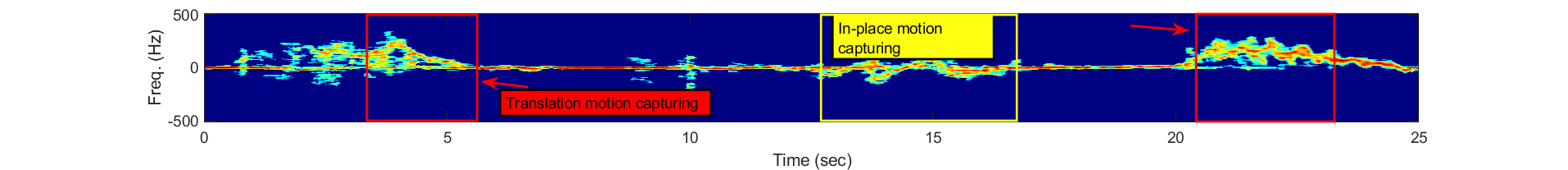}
\includegraphics[width=\linewidth]{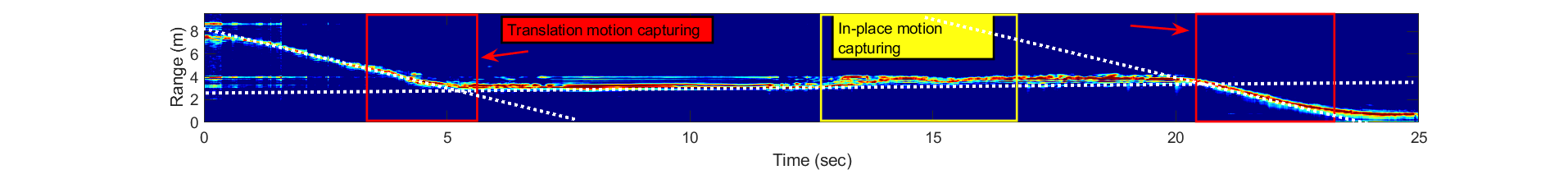}
\caption{Motion sequence: walking-falling, standing up from falling, walking. \label{fig:SequenceWlkFalSta}}
\end{figure*}
\begin{table}[htb]
\caption{First Post-Walking and last Pre-walking Motion Classification, [$d_{MD}=2$; $d_{RM}=1$].} 
\label{tab:clasMat_0a_mod}
\centering
\begin{tabular}{rrrrrrr}
\hline
 & \multicolumn{2}{c}{\begin{tabular}[c]{@{}c@{}}Micro-Doppler\\ Predicted\end{tabular}} & \multicolumn{2}{c}{\begin{tabular}[c]{@{}c@{}}Range-map\\ Predicted\end{tabular}} & \multicolumn{2}{c}{\begin{tabular}[c]{@{}c@{}}Fusion\\ Predicted\end{tabular}} \\ \hline
\multicolumn{1}{r|}{cl.} & (I) & \multicolumn{1}{r|}{(II)} & (I) & \multicolumn{1}{r|}{(II)} & (I) & (II) \\ \hline
\multicolumn{1}{r|}{(I)} & 98.8\% & \multicolumn{1}{r|}{1.2\%} & 96.6\% & \multicolumn{1}{r|}{3.4\%} & 100\% & 0\% \\
\multicolumn{1}{r|}{(II)} & 11.0\% & \multicolumn{1}{r|}{89.0\%} & 26.8\% & \multicolumn{1}{r|}{73.3\%} & 10.5\% & 89.5\%
\end{tabular}
\end{table}
\begin{table}[htb]
\caption{Post-walking in-place classification, [$d_{MD}=6$; $d_{RM}=2$]. } 
\label{tab:clasMat_1a_ext}
\scalebox{0.85}{
\begin{tabular}{rrrrrrrrr}
\hline
3 & \multicolumn{4}{c}{\begin{tabular}[c]{@{}c@{}}Micro-Doppler\\ Predicted\end{tabular}} & \multicolumn{4}{c}{\begin{tabular}[c]{@{}c@{}}Range-map\\ Predicted\end{tabular}} \\ \hline
\multicolumn{1}{r|}{cl.} & (I) & (II) & (III) & \multicolumn{1}{r|}{(IV)} & (I) & (II) & (III) & \multicolumn{1}{r}{(IV)} \\ \hline
\multicolumn{1}{r|}{(I)} & 99.6\% & 0.2\% & 0.0\% & \multicolumn{1}{r|}{0.2\%} & 98.9\% & 0.9\% & 0.1\% & \multicolumn{1}{r}{0.1\%} \\
\multicolumn{1}{r|}{(II)} & 0.2\% & 99.8\% & 0.0\% & \multicolumn{1}{r|}{0.0\%} & 1.1\% & 97.5\% & 0.0\% & \multicolumn{1}{r}{1.4\%} \\
\multicolumn{1}{r|}{(III)} & 0.6\% & 0.0\% & 99.4\% & \multicolumn{1}{r|}{0.0\%} & 1.3\% & 0.0\% & 98.7\% & \multicolumn{1}{r}{0.0\%} \\
\multicolumn{1}{r|}{(IV)} & 0.3\% & 0.1\% & 0.0\% & \multicolumn{1}{r|}{99.6\%} & 4.6\% & 5.6\% & 0.5\% & \multicolumn{1}{r}{89.3\%} \\ \cline{1-5}
\multicolumn{1}{r|}{1a} & \multicolumn{4}{c|}{Fusion Predicted} &  &  &  &  \\ \cline{1-5}
\multicolumn{1}{r|}{cl.} & (I) & (II) & (III) & \multicolumn{1}{r|}{(IV)} &  &  &  &  \\ \cline{1-5}
\multicolumn{1}{r|}{(I)} & 99.9\% & 0.1\% & 0.0\% & \multicolumn{1}{r|}{0.0\%} &  &  &  &  \\
\multicolumn{1}{r|}{(II)} & 0.1\% & 99.9\% & 0.0\% & \multicolumn{1}{r|}{0.0\%} &  &  &  &  \\
\multicolumn{1}{r|}{(III)} & 0.6\% & 0.0\% & 99.4\% & \multicolumn{1}{r|}{0.0\%} &  &  &  &  \\
\multicolumn{1}{r|}{(IV)} & 0.2\% & 0.8\% & 0.3\% & \multicolumn{1}{r|}{98.7\%} &  &  &  & 
\end{tabular}
}
\end{table}
In the example depicted Fig.~\ref{fig:SequenceWlkFalSta}, the Radon transform renders the intersection point at $t_1 \approx 5s$. Here, the PBC cannot separate the energies of the walking and falling motions. This merging between translation and the follow-on in-place motion can indicate a falling, a stopping, or a bending activity, each is in the direction of walking. The intersection point from the Radon transform typically occurs after the walking ceases. Accordingly, at $t_1$, a window of $2s$ is applied which captures $0.5s$ of the in-place and $1.5s$ of the translation motion (left red rectangulars). From the state diagram, and considering the transitions out of the WS, the employed classifier for the captured motion includes only two possible classes, namely, (\textsc{I}) walking-stopping \& walking-bending, leading to the StS, and (\textsc{II}) walking-falling, leading to the LS (Table~\ref{tab:clasMat_0a_mod}). It is clear from the Table~\ref{tab:clasMat_0a_mod} that there are no false alarms for falling, but there is a missing probability of $10.5\%$, which is high. If all motion classes are considered including those of the SiS, the missing probability of falling rises to $19.3\%$ (Table~\ref{tab:AllClassesFusion}), which is certainly unacceptable. The non-zero missing probability can mistakenly assign a StS instead of a LS for a fall. Therefore, all in-place motions from the StS and LS, facing the radar, should be considered for the next classifier. According to the state diagram, these motions are: (I) bending while standing, (II) sitting from standing, (III) falling from standing, and (IV) standing up from falling (Table~\ref{tab:clasMat_1a_ext}).

When applying the PBC, it is determined that the onset and offset times of the in-place motion are at $t_2 \approx 12.5s$ and $t_3 \approx 16s$, respectively. From the classification Table~\ref{tab:clasMat_1a_ext}, the ground truth motion (IV), which is standing up from falling, has a missing probability of $1.3\%$, whereas there is no probability of false alarm. In contrast, the missing probability when applying all classes is $3.4\%$.  Comparing Tables~\ref{tab:clasMat_0a_mod} and \ref{tab:clasMat_1a_ext}, it is evident that the detection of standing up from falling is more reliable than the detection of falling when it is closely merged with walking.

The classification outcome leads to the StS, which spreads until the intersection point of $t_4 \approx 20.5s$ before the Radon transform declares a WS. Over the window from $t_4$ to  ${t_4+3s}$, the classifier discriminates between the only two possible actions leading to the WS. According to the state diagram, these actions are: (I) starting-walking and (II) standing up  from sitting merged with  walking. Both motions are classified with $100\%$ accuracy (Table~\ref{tab:clasMat_4}).
%
\subsection{Pre-Walking Motion Classification} 
\begin{table}[htb]
\caption{First Pre-Walking Motion Classification and Last Post-walking Motion Classification, [$d_{MD}=14$; $d_{RM}=4$].} 
\label{tab:clasMat_4}
\centering
\begin{tabular}{rrrrrrr}
\hline
 & \multicolumn{2}{c}{\begin{tabular}[c]{@{}c@{}}Micro-Doppler\\ Predicted\end{tabular}} & \multicolumn{2}{c}{\begin{tabular}[c]{@{}c@{}}Range-map\\ Predicted\end{tabular}} & \multicolumn{2}{c}{\begin{tabular}[c]{@{}c@{}}Fusion\\ Predicted\end{tabular}} \\ \hline
\multicolumn{1}{r|}{cl.} & (I) & \multicolumn{1}{r|}{(II)} & (I) & \multicolumn{1}{r|}{(II)} & (I) & (II) \\ \hline
\multicolumn{1}{r|}{(I)} & 100\% & \multicolumn{1}{r|}{0\%} & 99.2\% & \multicolumn{1}{r|}{0.8\%} & 100\% & 0\% \\
\multicolumn{1}{r|}{(II)} & 0\% & \multicolumn{1}{r|}{100\%} & 0\% & \multicolumn{1}{r|}{100\%} & 0\% & 100\%
\end{tabular}
\end{table}
%
\begin{table}[htb]
\caption{Second Pre-walking Motion Classification, [$d_{MD}=7$;~$d_{RM}=2$].\label{tab:clasMat_R1}}
\centering
\scalebox{0.79}{
\begin{tabular}{@{}rrrrrrrrrr@{}}
\toprule
& \multicolumn{3}{c}{\begin{tabular}[c]{@{}c@{}}Micro-Doppler\\ Predicted\end{tabular}} & \multicolumn{3}{c}{\begin{tabular}[c]{@{}c@{}}Range-map\\ Predicted\end{tabular}} & \multicolumn{3}{c}{\begin{tabular}[c]{@{}c@{}}Fusion\\ Predicted\end{tabular}} \\ \midrule
\multicolumn{1}{r|}{cl.} & (I) & (II) & \multicolumn{1}{r|}{(III)} & (I) & (II) & \multicolumn{1}{r|}{(III)} & (I) & (II) & (III) \\ \midrule
\multicolumn{1}{r|}{(I)} & 98.7\% & 0.9\% & \multicolumn{1}{r|}{0.5\%} & 95.7\% & 4.1\% & \multicolumn{1}{r|}{0.3\%} & 98.9\% & 0.7\% & 0.4\% \\
\multicolumn{1}{r|}{(II)} & 0.6\% & 99.3\% & \multicolumn{1}{r|}{0.1\%} & 2.7\% & 97.2\% & \multicolumn{1}{r|}{0.1\%} & 0.7\% & 99.3\% & 0.0\% \\
\multicolumn{1}{r|}{(III)} & 0.5\% & 0.8\% & \multicolumn{1}{r|}{98.7\%} & 0.9\% & 4.2\% & \multicolumn{1}{r|}{94.9\%} & 0.0\% & 0.2\% & 99.8\%
\end{tabular}
}
\end{table}
With Table~\ref{tab:clasMat_4} classification certainty, we can go backward in time and revisit the in-place motion classified by Table~\ref{tab:clasMat_1a_ext} when the forward time motions are considered. If classifier \ref{tab:clasMat_4} declares a StS, then the actions occurring prior the StS would be  (I) standing up from sitting, (II) bending while standing or (III) standing up from falling. These actions originate from the SiS, the StS or the LS, respectively. The classification results are shown in Table~\ref{tab:clasMat_R1}. Standing-up from falling has only $0.2\%$ miss-detection probability and $0.4\%$ false alarm probability wrongly declaring bending while standing. 
These results in Table~\ref{tab:clasMat_4}+\ref{tab:clasMat_R1} are more assertive than those in Table~\ref{tab:clasMat_0a_mod}+\ref{tab:clasMat_1a_ext} of the forward in time motion classification. In this sense, the underlying example underscores the importance of considering both directions in rendering a decision.

\begin{table*}[hbt] 
\caption{Fusion classification rate for all ADL classes.}
\centering
\label{tab:AllClassesFusion}
\scalebox{0.92}{
\begin{tabular}{|l|rrrrrrrrrrrrrrrrr|} \hline
\rotatebox[origin=l]{-90}{~\diagbox[dir=SW]{\textbf{Pred. Classes}}{\textbf{True Classes}}~~}  & \rotatebox[origin=l]{-90}{\textbf{T-Walking-Stop/Bent (I)}~} & \rotatebox[origin=l]{-90}{\textbf{T-Walking-Fall (II)}~} & \rotatebox[origin=l]{-90}{A-Walking-Stop/Bent (III)~} & \rotatebox[origin=l]{-90}{A-Walking-Fall (IV)~} & \rotatebox[origin=l]{-90}{\textbf{T-Sitting down (V)}}  & \rotatebox[origin=l]{-90}{\textbf{T-Bending w. Standing (VI)}~} & \rotatebox[origin=l]{-90}{A-Bending w. Standing (VII)~} & \rotatebox[origin=l]{-90}{\textbf{T-Falling f. Standing (VIII)}~} & \rotatebox[origin=l]{-90}{A-Falling f. Standing (IX)~} & \rotatebox[origin=l]{-90}{\textbf{T-Standing f. Falling (X)}~} & \rotatebox[origin=l]{-90}{A-Standing f. Falling (XI)~} & \rotatebox[origin=l]{-90}{\textbf{T-Standing f. Sitting (XII)}~} & \rotatebox[origin=l]{-90}{T-Bending conn. f. Sitting (XIII)~} & \rotatebox[origin=l]{-90}{T-Bending down f. Sitting (XIV)~} & \rotatebox[origin=l]{-90}{T-Bending up f. Sitting (XV)~} & \rotatebox[origin=l]{-90}{\textbf{T-Standing up-Walking (XVI)}~} & \rotatebox[origin=l]{-90}{\textbf{T-Start Walking (XVII)}~}  \\   \hline \noalign{\vskip 0.05cm} 
\textbf{I} & 100.0\% & 0.0\% & 0.0\% & 0.0\% & 0.0\% & 0.0\% & 0.0\% & 0.0\% & 0.0\% & 0.0\% & 0.0\% & 0.0\% & 0.0\% & 0.0\% & 0.0\% & 0.0\% & 0.0\% \\
\textbf{II} & 15.9\% & 80.7\% & 0.0\% & 0.0\% & 0.0\% & 0.0\% & 0.0\% & 3.5\% & 0.0\% & 0.0\% & 0.0\% & 0.0\% & 0.0\% & 0.0\% & 0.0\% & 0.0\% & 0.0\% \\
\textbf{V} & 0.0\% & 0.0\% & 0.0\% & 0.0\% & 98.9\% & 0.0\% & 0.3\% & 0.0\% & 0.0\% & 0.0\% & 0.0\% & 0.0\% & 0.0\% & 0.0\% & 0.8\% & 0.0\% & 0.0\% \\
\textbf{VI} & 0.0\% & 0.0\% & 0.0\% & 0.0\% & 0.3\% & 94.0\% & 4.4\% & 0.0\% & 0.0\% & 0.0\% & 0.0\% & 0.2\% & 0.6\% & 0.6\% & 0.0\% & 0.0\% & 0.0\% \\
\textbf{VIII} & 0.0\% & 0.0\% & 0.0\% & 0.0\% & 0.0\% & 0.0\% & 0.0\% & 98.5\% & 0.0\% & 0.0\% & 0.0\% & 0.3\% & 0.0\% & 1.2\% & 0.0\% & 0.0\% & 0.0\% \\
\textbf{X} & 0.0\% & 0.0\% & 0.0\% & 0.0\% & 2.1\% & 0.0\% & 0.0\% & 0.0\% & 0.0\% & 96.6\% & 1.3\% & 0.0\% & 0.0\% & 0.0\% & 0.0\% & 0.0\% & 0.0\% \\
\textbf{XII} & 0.0\% & 0.0\% & 0.0\% & 0.0\% & 0.3\% & 0.1\% & 0.8\% & 0.0\% & 0.0\% & 0.0\% & 0.0\% & 95.3\% & 0.0\% & 3.4\% & 0.0\% & 0.1\% & 0.0\% \\
\textbf{XVI} & 0.0\% & 0.0\% & 0.0\% & 0.0\% & 0.0\% & 0.0\% & 0.0\% & 0.0\% & 0.0\% & 0.0\% & 0.0\% & 1.7\% & 0.0\% & 0.0\% & 0.0\% & 98.3\% & 0.0\% \\
\textbf{XVII} & 0.0\% & 0.0\% & 0.0\% & 0.0\% & 0.0\% & 0.0\% & 0.0\% & 0.0\% & 0.0\% & 0.0\% & 0.0\% & 0.0\% & 0.0\% & 0.0\% & 0.0\% & 0.0\% & 100.0 \\ \hline %
\end{tabular}
}
\end{table*}
\section{Conclusion\label{sec:conc}}
Excluding many or some of ADL from the decision on motion discrimination at any given time helps in improving classification results. The paper proposed forward and reverse time motion classifications, and showed, by example, that more reliable decision may be gleaned from looking backward in time. The classifier used is 2-D~PCA  data driven feature extraction where both the range-map and spectrogram micro-Doppler signatures are fused into one feature. We applied the Radon transform as a tracker, in essence, detecting the action of walking, or lack of, from the line behavior in the range-map. Intersection points revealed transitioning from translation to in-pace motion or vice versa.   



%

\balance
\bibliographystyle{IEEEtran.bst}
\bibliography{IEEEabrv,refs.bib}

\end{document}